\documentclass[sigconf,9pt,nonacm]{acmartmod}

\newif\ifEditMode
\EditModefalse

\usepackage{graphicx,tikz,amsmath,mathtools,tikz-cd,epstopdf,verbatim, hyperref}
\usepackage{wrapfig,xspace,caption,subcaption}
\usepackage[font=itshape]{quoting}
\usepackage{enumitem}
\usepackage[T1]{fontenc}
\usepackage{algorithm}
\usepackage[noend]{algpseudocode}
\usepackage{syntax,cjhebrew,proof}

\usetikzlibrary{automata, positioning, arrows}

\usepackage{amsthm}
\usepackage[breakable, theorems, skins]{tcolorbox}
\usepackage{stmaryrd}
\usepackage{booktabs,multirow}

\tikzset{
->, 
node distance=3cm, 
every state/.style={thick, fill=gray!10}, 
initial text=$ $, 
}

\PassOptionsToPackage{bookmarks=false}{hyperref}

\newtheorem{proposition}{Proposition}[section]

\theoremstyle{definition}

\newtheorem{example}{Example}[section]
\newtheorem{definition}{Definition}[section]

\graphicspath{{figures/}}

\allowdisplaybreaks

\ifEditMode
   \paperwidth=\dimexpr \paperwidth + 8cm\relax
   \oddsidemargin=\dimexpr\oddsidemargin + 4cm\relax
   \evensidemargin=\dimexpr\evensidemargin + 4cm\relax
   \marginparwidth=\dimexpr \marginparwidth + 4cm\relax
\fi

\DeclareRobustCommand{\mybox}[2][gray!20]{%
\begin{tcolorbox}[   
        breakable,
        left=0pt,
        right=0pt,
        top=-2pt,
        bottom=0pt,
        colback=#1,
        colframe=#1,
        width=\dimexpr\columnwidth\relax, 
        enlarge left by=0mm,
        boxsep=2pt,
        arc=0pt,outer arc=0pt,
        ]
        #2
\end{tcolorbox}
}

\setlist[enumerate]{leftmargin=*}
\setlist[itemize]{leftmargin=*}

\renewcommand{\models}{\mathrel{|}\joinrel\mkern-1.0mu\mathrel{=}}

\DeclareFontFamily{U}{musix}{}
\DeclareFontShape{U}{musix}{m}{n}{<-> s*[1.01] musix11}{}

\newcommand{\reals}[0]{\mathbb{R}}
\newcommand{\nnreals}[0]{\reals_{\ge 0}}
\newcommand{\natnums}[0]{\mathbb{N}}

\providecommand{\setunion}[0]{\boldsymbol{\cup}}
\providecommand{\setint}[0]{\boldsymbol{\cap}}

\newcommand{\clock}[0]{\mathcal{D}}
\newcommand{\clockspace}[0]{\mathfrak{C}}
\newcommand{\valspace}[0]{\mathcal{X}}
\newcommand{\ufunspace}[0]{\mathfrak{F}}
\newcommand{\lname}[0]{MCL\xspace}
\newcommand{\eventop}[0]{\lozenge}
\newcommand{\globop}[0]{\square}

\newcommand{\clockref}[0]{\rho}
\newcommand{\behunv}[0]{\mathbf{B}}
\newcommand{\cont}[0]{\mathcal{C}}

\providecommand{\setArg}[2]{\left\{ {#1} \,\,\middle|\,\, {#2} \right\}}
\providecommand{\cmp}[0]{M}

\providecommand{\reals}[0]{\mathbb{R}}
\providecommand{\fpoint}[0]{F}
\providecommand{\natnums}[0]{\mathbb{N}}
\providecommand{\nonneg}[0]{\reals_{\ge 0}}

\providecommand{\sat}[0]{\models}
\providecommand{\cmpComposition}[0]{\parallel}

\newcommand{\costfn}[0]{\mathcal{V}}

\newcommand{\varset}[0]{\mathbf{Vars}}
\newcommand{\clockpairs}[0]{\mathfrak{P}}

\begin{document}

\title{Specifying and Analyzing Networked and Layered Control Systems Operating on Multiple Clocks
}

\author{Inigo Incer}
\orcid{0000-0001-7933-692X}
\affiliation{%
  \institution{California Institute of Technology}
}

\author{Noel Csomay-Shanklin}
\orcid{0000-0002-2361-1694}
\affiliation{%
  \institution{California Institute of Technology}
}

\author{Aaron Ames}
\orcid{0000-0003-0848-3177}
\affiliation{%
  \institution{California Institute of Technology}
}

\author{Richard M. Murray}
\orcid{0000-0002-5785-7481}
\affiliation{%
  \institution{California Institute of Technology}
}

\renewcommand{\shortauthors}{Incer et al.}


\begin{abstract}

We consider the problem of reasoning about networked and layered control systems using assume-guarantee specifications.
As these systems are formed by the interconnection of components that operate under various clocks, we introduce a new logic, Multiclock Logic (MCL), to be able to express the requirements of components form the point of view of their local clocks. Specifying components locally promotes independent design and component reuse. We carry out a contract-based analysis of a control system implemented via two control algorithms (model predictive control and feedback linearization) running on their own processors and clocks. Then we implement each of the contracts to build a system. The system performs as desired when the requirements derived from our system-level analysis are respected. Violating the constraints required by the contract-based analysis of the system leads to error.
\end{abstract}

\maketitle

\section{Introduction}

Over the years, the field of control theory has developed many tools to design control blocks in isolation---e.g., PID, feedback linearization (FBL), model predictive control (MPC) \cite{allgower2004nonlinear,borrelli-predictive-nodate}, control Lyapunov functions (CLFs) \cite{artstein1983stabilization, sontag1989universal}, etc.
The design of complex control systems---legged robots, aerial robots, and autonomous vehicles, to name a few---normally involves the combination of various blocks of control functionality.
It is often the case that designers working independently on each control block make assumptions on the behaviors of other blocks that are not communicated or explicitly stated---leading to a development process prone to errors. 
Moreover, control blocks are normally implemented over various processors communicating over a network. The resulting communication/interfacing delays add imperfections that are often not modeled.
The end result is that we interconnect our control blocks \emph{hoping the system will work}.

It would help this situation if we had access to design tools that accepted descriptions of each control block together with constraints on communication delays, parameters uncertainties, etc., and which were able to judge whether the resulting system satisfies desired properties.
These tools could allow us to analyze the amount of imperfections our designs can tolerate while meeting our top-level objectives.
To be efficient, such tools should operate on an abstract version of each component, not on its detailed mathematical model. Using abstractions to represent components fosters component reuse.
To yield certainty that the system will work as intended, the tools should be deductive, not simulation-based.
In other words, to address current system-design hurdles, we envision formal, computer-aided design for system-level networked and layered control. As this is the undertaking of a major project, we focus on goals that can be addressed in the short term.

Our first objective towards specification-based analysis of control systems is the development of abstractions for control blocks.
We believe that control blocks should be abstracted into the specifications they satisfy, i.e., we should say what each component does, not how it does it. The specifications should be lighter-weight than the detailed mathematical models used to describe them.
Since our control blocks are open (i.e., they have inputs and outputs), these specifications should state what our components do and what they need in order to work. This form of specifications is precisely captured by \emph{assume-guarantee contracts}, which are formal specifications that state what a component guarantees, and what the component requires from its environment to be able to deliver its guarantees. Thus, we envision our control components specified using assume-guarantee contracts.
For instance, the assumptions of each component should include the freshness of the data required in order for them to carry out control-relevant tasks (e.g., a system won't work well if MPC uses old state estimates to compute trajectories, or if MPC takes too long to compute a feasible solution).
One benefit of using contracts is they come with a rich algebra that allows us to relate the specifications of the components with the specification of the entire system. Thus, we can ensure that the composition of the local contracts satisfies desired system-level objectives, such as stability and safety. We can also compute the specification of a component that we need to add to a system in order for our system to satisfy a top-level objective.
After we are satisfied with the system-level behavior of our system analyzed using contracts, we can implement each contract locally, ensuring that our components meet their specifications, and ensuring that each component has its assumptions satisfied. Our system level-analysis on specifications guarantees that the resulting system we implement will meet our desired objectives.

There is a major challenge we have to overcome in order to be able to specify general control systems.
Networked and multi-layer designs run processes on multiple processors which operate using multiple time bases, or clocks. 
In order to express properties of each control block in our design, we would like to do so locally, i.e., with respect to the time reference used by that particular block. This is key to enable independent design of control blocks. Currently, the control literature has used 
LTL \cite{PnueliLtl}, STL \cite{MalerSTL}, MTL \cite{Koymans1990}, and their variants to write specifications in numerous tasks. 
But these logics require us to define all variables with respect to the same time base, making them unsuitable for our objective that specifications should be written from a local perspective of a given timebase.
It would be ideal to have a logic that would let us write properties from the local point of view of each time base. This would enable i) compositional design and ii) separation of concerns: the engineers in charge of say, MPC, can develop their algorithms locally, making sure that their design provides certain guarantees as long as certain assumptions are satisfied, and system level analysis can show that a composition of such blocks will make the system behave as intended.

\textbf{Previous work.}
The use of formal assume-guarantee specifications has its origin in the pre/post conditions of Floyd-Hoare logic~\cite{floyd6719assigning,Hoare:1969}. The use of the word \emph{contract} was suggested by 
Meyer in the context of the language Eiffel \cite{meyerContract}.
Abadi and Lamport \cite{AbadiLamportComposingSpecs} proposed a principle to compose assume-guarantee specifications in order to link the specifications of components with the specification of the system.
The program of using assume-guarantee contracts to reason about any kind of system (not just software systems) appeared in the work of Benveniste et. al~\cite{multViewpoint,BenvenisteContractBook}. Incer's PhD thesis contains an analysis of all algebraic operations known on contracts and their relevance for system-level design \cite{incerphd}.

Contracts have been actively applied in controls.
Sangiovanni-Vincentelli et al.~\cite{sangiovanni2012taming} discuss a contract-based design methodology for cyber-physical systems.
Saoud et al.~\cite{SAOUD2021109910} analyze invariance properties on continuous-time dynamical systems using contracts.
Girard et al.~\cite{9993344} explore contracts and discrete-time dynamical systems and propose a link between invariant-set computation and the verification of satisfaction of contracts.
Liu et al.~\cite{liu2022compositional} recently considered the problem of synthesizing tasks described in signal temporal logic (STL) via contracts. Sharf et al.~\cite{SHARF202125} use contracts to reason about discrete-time dynamical control systems and discuss tools to verify contract satisfaction and refinement.
Phan-Minh and Murray~\cite{9654932} extend the theory of contracts to be able to specify reaction to system failures.

The intersection of contracts and control has recently found applications in the design of control protocols~\cite{Nuzzo14-1}, the design of the control of a glucose regulation system for patients of diabetes~\cite{9304434}, the compositional synthesis of vehicular mission plans~\cite{waqas2022contract}, traffic management~\cite{cai2023rules}, robotic mission planning~\cite{mallozzi2023contract}, the design of an autonomous aircraft taxi system~\cite{pinto2023leveraging}, and the design of space missions~\cite{pactispace}.

To the best of our knowledge, the problem of compositionally writing specifications for control systems made from components that operate on distinct time bases has not been studied. Similarly, the contract based analysis of these systems has not been pursued.


\textbf{Contributions.}
This paper proposes
a new logic called Multiclock Logic (\lname) that allows us to compositionally write properties for control systems from the point of view of any clock used in the system. This logic allows us to express properties of networked control systems.
We provide the syntax of MCL and define the behaviors on which we provide its semantics. 
We show how we can rigorously specify assume-guarantee contracts of multi-layer control systems using \lname using succinct specifications.
Using as an example the control of a pendulum implemented using MPC and feedback linearization, we show how to compute system-level contracts from the contracts of each control block. This allows us to conclude that our system has desired properties.

\smallskip
\textbf{Paper outline.}
Our paper is organized as follows.
In Section~\ref{sc:prelim} we review background knowledge on behavioral modeling and assume-guarantee contracts.
Section~\ref{sc:runningexampe} introduces our running example. We consider the control of a pendulum involving two control algorithms running on two processors: one computes desired trajectories, and the other makes the system follow those trajectories. This section describes how the design of such a system is normally done. We stress some aspects of the design process which today are often not modeled.
Section~\ref{sc:modeling} describes the mathematical framework that we use in order to reason about system executions involving multiple clocks.
Section~\ref{sc:mcl} introduces the syntax and semantics of a new logic introduced in this paper, Multiclock Logic (\lname), whose purpose is to facilitate the expression of properties with respect to various timing references.
Section~\ref{sc:appl} discusses the specification of the control components from our running example. We also carry out system-level verification using the component contracts we express. We develop implementations for each control block and show that the satisfaction of system-level constraints leads to systems that meet our desired fuctionality, while violating requirements leads to failure.

\section{Preliminaries}
\label{sc:prelim}

This section provides background on behavioral modeling and assume-guarantee contracts, both of which are used throughout the paper.

\subsection{Behavioral modeling}
\label{sc:behmodel}

{\color{black}
Behavioral modeling understands components as the set of behaviors they can display. A behavior is a formalized execution of a component, an instance of the component operating. 
To illustrate the notion of a behavior, consider the voltage amplifier $M$ shown in Figure~\ref{byoiy98y}. The device has input $x$ and output $y$. Suppose we are only interested in the static operation of the amplifier. In that case, the notion of time is not needed in our description. If we had a perfect amplifier whose output is exactly equal to its input, we could say that the behaviors are all pairs of real numbers $(x,y)$ such that $y = x$. In this case, each of these pairs is a behavior. Formally, we would express the amplifier as
$\cmp = \setArg{(x,y) \in \reals^2}{y = x}$.

If we were interested in expressing dynamic attributes of a component, i.e., how it changes over time, the behaviors we just used would be insufficient. Now we would need to tell how the inputs and outputs, and possibly state variables, vary over time. If we assume that time is a continuous variable that takes values in the nonnegative real numbers, $\nonneg$, then each behavior of the amplifier could be expressed as a function with domain $\nonneg$ and codomain $\reals^2$ (for the variables $x$ and $y$). We could write the amplifier as
$$\cmp = \setArg{f \in \nonneg \to \reals^2}{f_1(t) = f_2(t) \text{ for all } t \in \nonneg},$$
where the subscript notation means the projection of the codomain $\reals^2$ to the corresponding component (we assume the first component is $x$, and the second $y$).

The examples we considered are continuous, but we can also behaviorally model discrete systems, such as $\cmp''$ of Figure~\ref{byoiy98y}. Here the indicated variables could be thought of as floating point numbers, and the operations as arithmetic over these numbers; the registers capture the state of the variables at certain times. $\cmp''$ can be written as
{\footnotesize
$$
M '' = \setArg{f \in \natnums \to \fpoint^4}{ \text{ For all } t \in \natnums,\,\, \begin{aligned} & f_3(t + 1) = f_1(t) + f_2(t) \text{ and} \\ & f_4(t+2) = f_3(t+1) f_2(t) \end{aligned} },
$$}
where $\fpoint$ is the set of floating-point numbers.
\begin{figure}
    \centering
    \includegraphics[width=0.8\columnwidth]{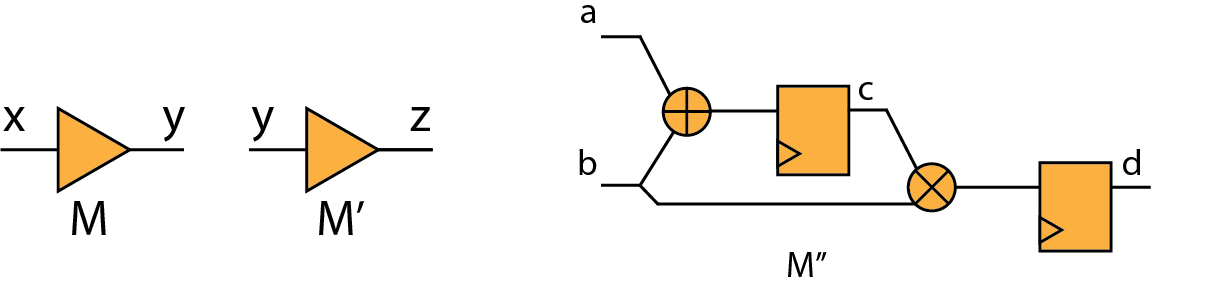}
    \caption{Examples of engineering systems to be modeled behaviorally}
    \label{byoiy98y}
    \vspace{-3mm}
\end{figure}

The functional notation for the representation of behaviors allows us to abstract whether the behaviors are continuous or discrete, whether their domain is bounded or not. We now introduce our formal definition of components and properties:

\begin{definition}
Let $\behunv$ be a set whose elements we call behaviors.
A \emph{component} is a subset of $\behunv$. Similarly, a \emph{property} is a subset of $\behunv$.
\end{definition}

By defining $\behunv$, we assume that we have already decided on the formalism we use to model our components and the types of properties we wish to verify for those components.
The difference between a component and a property is how we use them: we think of components as sets containing the behaviors that a design entity can display, and we understand a property as the collection of behaviors having a quality of interest, like safety or liveness. When we speak of a component having a property, we mean this:

\begin{definition}
\label{lknhbkgqb}
Suppose $\cmp$ is a component, and $P$ a property. We say $\cmp$ \emph{satisfies} $P$, written $\cmp \sat P$, when
$M \subseteq P$.
\end{definition}
In other words, $\cmp$ satisfies the property when all behaviors of $\cmp$ have a quality of interest.
Our next definition pertains the construction of systems by interconnecting components.
\begin{definition}
\label{qkjbgdjqhwg}
Let $\cmp$ and $\cmp'$ be components defined over a set $\behunv$. The \emph{composite} of $\cmp$ and $\cmp'$, denoted $\cmp \cmpComposition \cmp'$, is the component $M \cmpComposition \cmp' = M \setint \cmp'$.
\end{definition}
If we interpret $\cmp$ as the behaviors satisfying a certain constraint, and $\cmp'$ as those satisfying another, the composite is the component containing the behaviors satisfying both constraints simultaneously. This notion of composition is independent of the topology of the connection between $\cmp$ and $\cmp'$; the topology is implicitly included in the collections of behaviors of each component.

As an example, consider components $\cmp$ and $\cmp'$ from Figure~\ref{byoiy98y}. When we discuss the composition of components, it is necessary that all their behaviors be defined with respect to the same variables, that is, both $\cmp$ and $\cmp'$ must refer to variables $x, y, z$. Thus, we could express components $M$ and $M'$ as follows:
\begin{align*}
    M  &= \setArg{f \in \nonneg \to \reals^3}{f_1(t) = f_2(t) \text{ for all } t \in \nonneg} \\
    M' &= \setArg{f \in \nonneg \to \reals^3}{f_2(t) = f_3(t) \text{ for all } t \in \nonneg}.
\end{align*}

Observe that $M$ imposes no restrictions on the third component (corresponding to variable $z$), and $M'$ imposes no restrictions on the first (corresponding to variable $x$). The composition of these two objects is
\begin{align*}
M \cmpComposition M' &= \setArg{f \in \nonneg \to \reals^3}{
\begin{aligned}
&f_1(t) = f_2(t) \text{ and } f_2(t) = f_3(t) \\
&\text{ for all } t \in \nonneg
\end{aligned}    
},
\end{align*}
which corresponds to our understanding that $z = y = x$ when the two components are used concurrently. Observe that the variables used by the components determine the connection between them, i.e. because the output of $M$ is $y$ and the input of $M'$ is $y$, their composition connects the output of $M$ to the input of $M'$.

}

\subsection{Assume-guarantee contracts}

One of our objectives is to abstract control elements into local specifications and carry out system-level analysis using these abstractions.
When we build a system implementation, we want the result of our analysis to hold for our system implementation as long as the implementation of each system component adheres to its local specification.
This is the essence of compositional design: to separate system-level analysis from the implementation details of a given component in the system.
The theory of assume-guarantee contracts was designed to explicitly support these use cases. Throughout this paper, we will write specifications expressed as assume-guarantee (AG) contracts.

AG contracts \cite{BenvenisteContractBook,incerphd} are formal specifications that state i) assumptions made on the environment in which the component under specification operates and ii) the guarantees that the device under specification will provide when the assumptions of the contract are satisfied. In this way, AG contracts are akin to component datasheets.
This definition makes contracts suitable tools for the specification of open systems. As open systems interact with their environments, it is natural that they should state what they expect from their environments in order to work, and what they promise when the environment meets their assumptions.

A contract $\cont$ can be expressed as $\cont = (a,g)$, where $a$ is a formula in a suitable logic representing the \emph{assumptions} and $g$ a formula representing the \emph{guarantees} of the contract.
The type of expression that we use to express assumptions and guarantees depends on the underlying logic being used, e.g., LTL, STL, etc.
Since contracts are formal specifications, we need a way to relate contracts to their implementations.

\begin{definition}
\label{kgfqkgb}
Let $\behunv$ be a universe of behaviors.
We say that a component $\cmp$ defined over $\behunv$ is an \emph{implementation} of $\cont$, denoted $\cmp \sat \cont$, if $\cmp \models a \to g$, i.e., if $M$ provides the guarantees of the contract as long as the assumptions hold.
\end{definition}
Note that the verification of the notion of satisfaction indicated by the symbol $\models$ is carried out depending on the way in which components and formulas are expressed.
Definition~\ref{kgfqkgb} allows us to relate components with their specifications. In order to determine when a contract is stricter or more demanding than another, we use the notion of \emph{refinement}:
\begin{definition}
Let $\cont = (a,g)$ and $\cont' = (a', g')$ be two contracts. We say that $\cont \le \cont'$, i.e., that $\cont$ is a \emph{refinement} of $\cont'$, if
$a' \le a$ and $a \to g \le a' \to g$.
\end{definition}
For formulas, the notation $f \le f'$ means that $f \to f'$ is a tautology.
Refinement allows us to compare contracts. The next notion we need is that of combining contracts. One of our purposes is to replace the notion of component composition with the notion of contract composition. This is key, as it will allows us to carry out system level analysis using contracts. We will state as a definition an operation that is usually derived from a high-level axiom.

\begin{definition}
Let $\cont = (a,g)$ and $\cont' = (a', g')$ be two contracts. Their \emph{composition} is
$$\cont \parallel \cont' = \left( (a \land a') \lor (a \land \neg g) \lor (a' \land \neg g') , (a \to g)\land(a' \to g') \right).$$
\end{definition}
The composition of contracts yields the specification of the system obtained by composing implementations of the contracts being composed.
The key result that enables specification-based design is the following:
\begin{proposition}\label{kbgfkjqgf}
Let $\cont$, $\cont'$, and $\cont''$ be contracts and let $\cmp$ and $\cmp'$ be implementations of $\cont$ and $\cont'$, respectively. Suppose that $\cont \parallel \cont' \le \cont''$. Then $\cmp \parallel \cmp' \sat \cont''$.
\end{proposition}
This proposition says that if the contract composition $\cont \parallel \cont'$ refines $\cont''$, then
$\cmp \parallel \cmp' \models \cont''$. This means that we can carry out system analysis at the level of specifications, and we know in advance that the result of our analysis will hold when we build the system using implementations of each contract.

\section{Running example: two layers of a multi-rate system}
\label{sc:runningexampe}

\begin{figure}
\centering
\includegraphics[width=0.95\columnwidth]{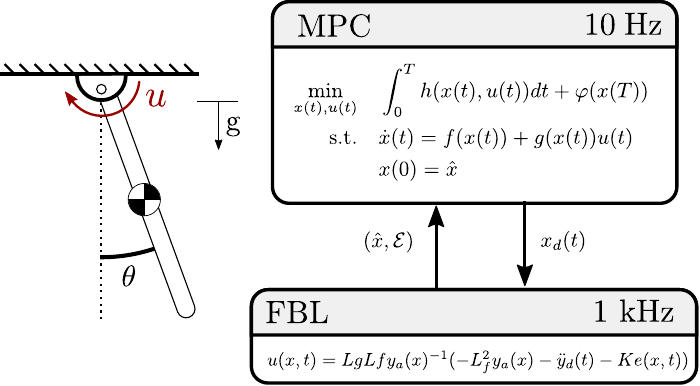}
\caption{(a) The nonlinear system considered in our case study. (b) Architecture of the hierarchical control system.}
\label{fig:exampleSystem}
\vspace{-2mm}
\end{figure}

Our objective is to carry out a system-level analysis of networked and layered control systems using specifications. To illustrate these ideas, we will consider the design of a control strategy for the pendulum show in in Figure~\ref{fig:exampleSystem}a. We will demonstrate that violations of the assumptions made in the development of the control hierarchy, shown in Figure~\ref{fig:exampleSystem}b, will lead to constraint violation of the closed-loop control system and ultimately to system failure. To begin, the configuration space of the system is given by $\theta\in\mathcal{Q}\triangleq \mathbb{S}^1$ and the associated state by $x=(\theta, \dot \theta)\in \mathcal{X}\triangleq\mathsf T \mathcal{Q}$. We can write the dynamics in control-affine form as
\begin{align*}
    \dot x = f(x) + g(x) u,
\end{align*}
with control input $u\in \mathbb{R}$, continuously differentiable drift vector $f:\mathcal{X} \to \mathbb{R}^2$, and actuation matrix $g:\mathcal{X}\to\mathbb{R}^2$.
We will be concerned with the high level task of reaching some neighborhood of the goal state $x_g = 0$, the unstable upright equilibrium. Low-level controllers alone struggle to simultaneously enforce state and input constraints while ensuring progress to the goal is being made \cite{borrelli-predictive-nodate}---%
this motivates the use of a hierarchy for achieving this control task. Our hierarchy will consist of two control blocks, a model-predictive controller and a feedback linearization layer.

\subsection{High-level controller design---MPC}
The MPC controller will be concerned with producing the pieces of trajectories $x_d$ which make progress towards the goal $x_g$ for the low-level controller to track. In order to do so, we set up the following optimal control problem:

{\footnotesize
\begin{equation}
\begin{aligned}
\min_{x(t), u(t)} \quad & \int_0^T h(x(t), u(t)) dt + \varphi(x(T)) \\
\textrm{s.t.} \quad & \dot x(t) = f(x(t)) + g(x(t))u(t) \\
  & x(t) \in \mathcal{X},~~ u(t) \in \mathcal{U}\\
  &x(0) = \hat{x}
\end{aligned}
\end{equation}}%
where $\hat x\in \mathcal{X}$ denotes an estimate of the system state; $h\colon \mathcal{X}\times \mathbb{R} \to \mathbb{R}$ is the running cost; and $\varphi\colon \mathcal{X}\to\mathbb{R}$ is the terminal cost. Solving this optimal control problem in a receding-horizon fashion, i.e., replanning $x_d$ at each clock of the high level, is termed MPC. As this optimization program is non-convex and infinite dimensional, often overlooked approximations such as convexifying the dynamics and discretizing the state-input trajectory must be made in order to make solving this problem computationally tractable.

\subsection{Low-level controller design}
To actuate the system, we use a feedback controller which is able to track a desired trajectory $x_d$ received from the high level controller. We begin by defining an output $y\colon\mathcal{X} \to\mathbb{R}$ as $y(x) = \theta$, whereby we can construct error coordinates
\begin{align*}
    e(x, t) = \underbrace{\begin{bmatrix}
        y(x) \\ L_fy(x)
    \end{bmatrix}}_{\eta(x)}-\underbrace{\begin{bmatrix}
        y(x_d(t)) \\ \dot y(x_d(t))
    \end{bmatrix}}_{\eta_d(t)}, 
\end{align*}
with $L_fy(x)$ denoting the Lie derivative of $y$ with respect to $f$. Then, a feedback control law can be created to stabilize the outputs. For example, the control law can be the feedback linearizing controller
\begin{align}
    k_\text{fbl}(x, t) = L_gL_f(x)^{-1}(-L_f^2y(x) + \ddot y(x_d(t)) - Ke(x, t)),
    \label{eq:fbl}
\end{align}
which exponentially stabilizes the output coordinates for elementwise positive vector $K\in\mathbb{R}^{2}$. 

Combining MPC with a low-level controller is a popular technique technique which is extensible to stabilizing complex nonlinear systems \cite{grandia2022perceptive, rosoliaunified2022, csomay-shanklinnonlinear2023}. When deploying such an architecture, however, the underlying approximations are often not explicitly reasoned about, leaving practitioners hoping rather than knowing that the closed-loop system will work. Common assumptions include:
\begin{itemize}
    \item The estimate of the state $\hat x$ is ``close enough'' to the true state $x$.
    \item The MPC runs ``fast enough'' and produces solutions which are ``close enough'' to the system dynamics.
    \item The low-level controller runs ``fast enough" and can track ``well enough.''
\end{itemize}
The aim of this paper is to construct a pipeline whereby these assumptions are made explicit, with an eye towards automating the system verification process for control systems. We will later show examples of this verification process, and demonstrate instances when the closed-loop system will fail when these assumptions are violated. Before doing this, we will introduce i) our behavioral modeling framework to reason about networked/layered control scenarios and ii) a new logic that will allow us to succinctly express requirements across multiple clocks in the system.

\section{Behavioral modeling for networked and layered control}
\label{sc:modeling}

This paper introduces a logical formalism in which we can express and manipulate properties of relevance to networked control---the focus of Section~\ref{sc:mcl}.
This logical formalism must make predicates over well-defined system executions.
The purpose of this section is to apply the behavioral modeling introduced in Section~\ref{sc:behmodel} to systems involving networked and layered control. Its main contribution is summarized in Definition~\ref{kqbfgfbg}, which provides the universe of behaviors $\behunv$ that we will use to reason about control systems spanning multiple processors and clocks.
Once we have access to $\behunv$, the development of
Section~\ref{sc:behmodel} gives us access to
components and properties (Definition~\ref{lknhbkgqb}) and to component composition (Definition~\ref{qkjbgdjqhwg}).
The logic introduced in Section~\ref{sc:mcl} will be given a denotation based on the concepts introduced in this section.

\subsection{Variables and behaviors}

The most basic concept in system modeling is \emph{the variable}. A variable is a named entity in our system having a progression of valuations. The notion of progression is given by an ordered structure, and valuations are provided by a topological space.
\begin{definition}
A \emph{variable} is a tuple
$
    (v, \valspace_v, \clock_v)
$,
where $v$ is the \emph{name} for the variable,
$\valspace_v$ is a topological space called the \emph{value space},
and $\clock_v$ is a totally-ordered group called the \emph{clock}, giving a notion of a progression.
\end{definition}

We require the clock to be an ordered structure to be able to have semantic support for intervals, as we will see in Section~\ref{sc:mcl}. We also require the clock to be a group. We recall that
\begin{definition}
A group is a structure $(G, +, -, 0)$, where $+$ is an associative binary operation on $G$, $0 \in G$ is the identity of this operation, and $-$ is a unary operation such that for any $x \in G$, $x + (-x) = 0$.
\end{definition}
We will have the need to add and subtract elements of the clock structures. This is why we add group structure to the clocks.
We will use interchangeably the variable name and the tuple.
Observe that, according to our definition, a variable is a \emph{name}.
To speak about the values taken by a variable, we refer to its behaviors.

\begin{definition}
A \emph{behavior} of a variable $(v, \valspace_v, \clock_v)$ is a member of the set $\valspace_v^{\clock_v}$, i.e., the set of functions from $\clock_v$ to $\valspace_v$.
\end{definition}

\begin{example}
Suppose $S$ is a variable denoting a temperature in our system. We assume that temperature can be represented by any real number. If we assume that $S$ can take values for any real number, the tuple representing this variable is
$(S, \reals, \reals)$.
\end{example}

\begin{example}
Suppose $A$ is a variable used to model the behavior of a label in a state machine. The valuations of $A$ are Boolean, and $A$ takes values over a countable sequence. Thus, this variable is given by
$(A, \{0, 1\}, \natnums)$.
\end{example}

\begin{example}
    Suppose $D$ is a variable used to model a constant in the system. The valuations of $D$ are real. This variable is given by
    $(D, \reals, \{\bullet\})$.
\end{example}

\subsection{Multiple clocks}

The behaviors of a variable are given with respect to its clock.
When we consider multiple variables simultaneously in a system,
we will want to regard some of these variables as sharing the same clock. We will assume we have a set $\clockspace$ of clocks. Suppose a set $\varset$ of variables shares clock $c \in \clockspace$. Then the joint behaviors of these variables will be given by maps
$$
c \to \prod_{v \in \varset} \valspace_v.
$$
In order to isolate the value of a variable $u \in \varset$ from those of the other variables in the same clock, we will use the projection maps that come with the definition of the product:
$$
\begin{tikzcd}
        \prod_{v \in \varset} \valspace_v \arrow[r, "\pi_u"]
        & \valspace_u.
\end{tikzcd}
$$

\subsection{Relating clocks}

In order to state certain predicates,
we may need to use a clock to read the values of variables belonging to another clock.
For example, to state the accuracy of a digital temperature sensor, we may need to relate the digital value read by the sensor to the physical value of temperature clocked by a continuous clock.

Suppose $c, d \in \clockspace$. To read the value of a $d$-variable in the clock $c$, we will assume the existence of a map

{
\begin{center}
\begin{tabular}{c }
    $\begin{tikzcd}
        c \arrow[r, "\tau_c^{d}"]
        & d.
    \end{tikzcd}$
\end{tabular}
\end{center}
}

The $\tau$ maps, which we may call \emph{synchronization maps}, determine the index of the target clock that is used when reading variables from a given clock. Suppose, for example, that variable $x$ is clocked by clock $d$. If we want to read the value of $x$ from clock $c$ at time $t \in c$, we would get the value of $x$ at time $\tau_c^d (t)$. The synchronization maps capture the amount of delay in which we incur by reading data from a clock from which the data does not originate. Thus, bounding network delays in our systems boils down to placing constraints on the synchronization maps.

\subsection{System executions and components}

Suppose that we build a system that comprises a set of variables $\varset$ and a set of clocks $\clockspace$. For each variable $v$, we let $C(v)$ be the clock corresponding to $v \in \varset$. The inverse of this map, $C^{-1}(c)$, yields the set of variables corresponding to clock $c \in \clockspace$.

\begin{definition}
\label{kqbfgfbg}
The \emph{universe of system behaviors} over $\clockspace$ and $\varset$ is defined as
\[
\behunv = \prod_{d \in \clockspace} \left( d \to \prod_{v \in C^{-1}(d)} \valspace_v \right) \times
\prod_{c,d \in \clockspace} (c \to d).
\]
A \emph{behavior}, or execution, of the system is an element of $\behunv$.
\end{definition}

Thus, we can express a system behavior $\beta$ as
$$\beta = \left(b_d\right)_{d \in \clockspace} \times \left(\tau_c^{d}\right)_{c,d \in \clockspace}.$$
The elements $b_d$ of the tuple carry the behaviors of the variables clocked by $d$, while the $\tau_c^{d}$ indicate how to read values of $d$-variables using the clock $c$.

Now that we have the notion of a universe of behaviors
$\behunv$ over a clock set $\clockspace$ and variable set $\varset$, 
components and properties are given by Definition~\ref{lknhbkgqb}, and the
composition of components to build systems is given by Definition~\ref{qkjbgdjqhwg}.
Finally, we provide an additional definition that will be useful to give semantics to the eventual modality $\eventop$ of the logic.


\begin{definition}\label{xjhgkjhbd}
Let $\beta = \left(b_d\right)_{d \in \clockspace} \times \left(\tau_d^{d'}\right)_{d,d' \in \clockspace}$ be a system behavior. 
Given $t \in c \in \clockspace$, we define the $(c,t)$-execution $\beta_c^t$ as
$\beta_c^t = \left(b_d\right)_{d \in \clockspace} \times \left(\tilde \tau_d^{d'}\right)_{d,d' \in \clockspace}$, where $\tilde \tau_d^{d'} (x) = \begin{cases}\tau_{c}^{d'} (x + t) & d = c\\ \tau_{d}^{d'} (x) & d \ne c  \end{cases}$.
\end{definition}

The only difference between $\beta$ and $\beta_c^t$ is that in
$\beta_c^t$ the clock $c$ is anticipated by $t$ units.

\section{Multi-clock logic}
\label{sc:mcl}

Section~\ref{sc:modeling} introduced a universe of behaviors that we can use to reason behaviorally about networked and layered control systems. In this section, we introduce a new logical system, Multi-clock Logic (MCL), in which we can state propositions with respect to a specified timing reference, or clock, in our system.

\subsection{Syntax}

Assume we have access to a set $\varset$ of variables, a set $\clockspace$ of clocks, and a set $\ufunspace$ of formulas of various arities.
We let $\clockpairs$ be the set of symbols $d^{d'}$, where $d$ and $d'$ denote clocks.
The syntax of \lname is
\begin{align*}
\phi &::= \clockref c. \; \Phi \;|\;
          \neg \phi \;|\;
          \phi \land \psi \\
\Phi &::= P\left(\lambda_1(t_1), \ldots, \lambda_n(t_n)\right) \;|\;
          \neg \Phi \;|\;
          \Phi \land \Psi \;|\;
          \eventop_{[t_1,t_2]} \Phi \;|\;
          \eventop_{t_1} \Phi
          ,
\end{align*}
where $c \in \clockspace$,
$\lambda_i \in \varset \setunion \clockpairs \setunion \clockspace$, $t_i \in c$, and $P \in \ufunspace$ is an $n$-ary formula. 

We think of the formulas $\phi$ as the global syntax of \lname, and of the $\Phi$ formulas as the local syntax that applies to a clock.
The global syntax supports propositional logic.
The syntax $\phi = \clockref c. \; \Phi$ indicates that a global formula $\phi$ is created by binding a local formula $\Phi$ to the clock $c$.
The local syntax supports propositional logic and the eventual modality. It also supports the enunciation of predicates involving multiple variables evaluated at the indicated times $t_i \in c$.
The syntax also supports the enunciation of predicates on clocks $c \in \clockspace$ and on ``clocks with respect to other clocks'' $c^d \in \clockpairs$. The use of clocks and \emph{clock pairs} $c^d \in \clockpairs$ in formulas allows us to express timing constraints in our systems.
Observe that the logic leaves abstract the formulas used to define the predicates. The formulas are functions that evaluate to Boolean values.

\subsection{Semantics}

The semantics of an \lname formula apply to a system behavior. Let $\beta = \left(b_d\right)_{d \in \clockspace} \times \left(\tau_d^{d'}\right)_{d,d' \in \clockspace}$ be a system behavior. We have the following global semantics:
\begin{itemize}
    \item $\beta \models \clockref c. \; \Phi$ iff $\beta \models_c \Phi$
    \item $\beta \models \neg \phi$ iff $\beta \not \models \phi$
    \item $\beta \models \phi \land \psi$ iff $\beta \models \phi$ and $\beta \models \psi$
\end{itemize}
Formulas at the global level are either created by Boolean connectives, or they are local formulas that are assigned to a clock in the system. The symbol ``$\models_c$'' stands for satisfaction in the local semantics of \lname, defined as follows:
\begin{itemize}
    \item $\beta \models_c P\left(\lambda_1(t_1), \ldots, \lambda_n(t_n)\right)$ iff $P \left( \texttt{Interp}(\lambda_i, t_i) \right)_{i = 1}^n$, where 
    $$\texttt{Interp}(\lambda_i, t_i) = \begin{cases}
        \pi_{v} \circ b_{C(v)} \circ \tau_{c}^{C(v)} (t_i) & \lambda_i = v  \in \varset \\
        \tau_c^d (t_i) & \lambda_i = d \in \clockspace \\
        \tau_{d}^{d'} \circ \tau_{c}^d (t_i) & \lambda_i = d^{d'} \in \clockpairs \text{ and } d \ne c
    \end{cases}$$
    \item $\beta \models_c \neg \Phi$ iff $\beta \not \models_c \Phi$
    \item $\beta \models_c \Phi \land \Psi$ iff $\beta \models_c \Phi$ and $\beta \models_c \Psi$
    \item $\beta \models_c \eventop_{[t_1,t_2]} \Phi$ iff $\exists t .\; (t_1 \le t \le t_2) \land \left(\beta_c^t \models_c \Phi\right)$
    \item $\beta \models_c \eventop_{t_1} \Phi$ iff $\exists t .\; (t_1 \le t) \land \left(\beta_c^t \models_c \Phi\right)$
\end{itemize}
In addition to the formation of local formulas using Boolean connectives and the eventual modality $\eventop$, \lname local formulas can be formed by making a predicate over a set of symbols $\lambda_i$ and a set of values $t_i$ of the clock $c$. In general, the meaning of the formula $P\left(\lambda_1(t_1), \ldots, \lambda_n(t_n)\right)$ is the evaluation of the $n$-ary formula $P$
for the values that each symbol $\lambda_i$ takes when the value of the clock $c$ is $\tau_c^c(t_i)$. For example, the predicate
$\clockref c.\; \|x(0)\| > 2$ for a variable $x$ means that we will evaluate the value of $x$ when $c = \tau_c^c(0)$. If $x$ is clocked by $c$, we simply extract the value of $x$ at time $\tau_{c}^{c} (0)$; if it is clocked by $d \ne c$, we would return its value at time $\tau_c^d (0)$.
We recall that the local semantics can anticipate the value of $c$ by $t$ time units---see Definition~\ref{xjhgkjhbd}.

The formula $\clockref c.\; d(0) - K > T$ for $d \in \clockspace$ and constants $T, K \in d$ has the semantics
$\tau_c^d (0) - K > T$. 
This statement is true at ticks of clock $c$ for which the index of clock $d$ that $c$ uses to read the data of $d$ is larger than $K + T$.
This type of predicate can be used as a precondition that ensures that the clock $d$ has issued sufficient ticks.
The semantics of a formula of the form $\clockref c.\; r(0) - d^r(0) < T$ for clocks $c, d, r$,
is $\tau_c^r(0) - \tau_d^r \circ \tau_c^d (0) < T$.
It states that the difference in $r$ time units between a tick of $c$ and the tick of $d$ from which $c$ reads the data of $d$ must be less than $T$. This kind of statement is useful to bound the maximum allowed time for data to have existed in its local clock $d$ before it is read by another clock $c$. We will make use of this statement in our applications described in Section~\ref{sc:appl}.

Finally, the notion of satisfaction is extended from system behaviors to components as follows:
we say that a component $M$ satisfies an \lname formula $\phi$ if $\beta \models \phi$ for all $\beta \in M$.

\subsection{Extended syntax}

We extend the given syntax and semantics in the following standard ways. We will make use of the rest of the Boolean connectives $\lor$ and $\rightarrow$ defined in the usual way to express both global and local \lname formulas. We also introduce a second modality $\globop$ for ``globally'':
\begin{itemize}
    \item $\beta \models_c \globop_{[t_1,t_2]} \Phi$ iff $\beta \models_c \neg \eventop_{[t_1,t_2]} \neg \Phi$
    \item $\beta \models_c \globop_{t_1} \Phi$ iff $\beta \models_c \neg \eventop_{t_1} \neg \Phi$
\end{itemize}

Finally, when stating a formula of the form $\clockref c.\; P\left(\lambda_i(t_i) \right)_i$, we will sometimes omit the $t_i$ arguments. In that case, the parameter should be understood as being equal to $0$.

\section{Specifying and verifying the running example}
\label{sc:appl}

\newcommand{\mclock}[0]{\mathscr{m}}
\newcommand{\lclock}[0]{\mathscr{l}}
\newcommand{\rclock}[0]{\mathscr{r}}
\newcommand{\counter}[0]{\texttt{cnt}}
\newcommand{\upd}[0]{\texttt{upd}}

\newcommand{\closepred}[0]{\texttt{Close}}
\newcommand{\ball}[0]{\texttt{Ball}}
\newcommand{\succop}[0]{\texttt{Succ}}

In this section, we retake the analysis of the control system described in Section~\ref{sc:runningexampe}. Our objective is to specify the MPC and low-level control blocks independently and carry out system-level analysis using these specifications. The objective of our system-level analysis is to show that the system reaches and remains inside a neighborhood of the goal state $x_g$. To do so, we assume we have a cost function $\costfn$ that maps the state $x$ of the system to a well-ordered set. The value of $\costfn$ is zero in a neighborhood of $x_g$. The function $\costfn$ will be further specified when we consider the implementations of component contracts in Section~\ref{jbhkxnlbbg}. 
As a well-order does not have infinite descending chains, the fact that the codomain of $\costfn$ is a well-order means that any process that decreases the cost $\costfn$ will eventually reach the minimum cost $\costfn(x) = 0$ in a finite number of iterations.

The system has three clocks: $\mclock = \natnums$, which runs the MPC block; $\lclock = \natnums$, which runs the low-level and estimation blocks; and $\rclock = \nnreals$, the physical time.
The system state is denoted by the variable $(x, \mathcal{X}, \rclock)$, i.e., its behaviors are functions from the physical clock to the state space $\mathcal{X}$.
The MPC block outputs a variable $(x_d, \mathcal{X}^\rclock,\mclock)$, which contains the trajectory in the state space that the system is to follow. Observe that the behaviors of this variable are functions from $\mclock$ to functions from $\rclock$ to $\mathcal{X}$. Thus, at any tick of the clock $\mclock$, MPC will provide a function for the low-level controller. This means that $x_d$ will be doubly-indexed in our formulas. The first index corresponds to the clock evaluation, and the second to the time argument of the trajectory. For example, the $\lclock$ formula $\clockref \lclock.\; \| x_d(0)(T) \| < K$ evaluates to
$\left\| x_d\left(\tau_\lclock^\mclock (0) \right)(T) \right\| < K$. Finally, an estimation block running on clock $\lclock$ will output a state estimate $(\hat x, \mathcal{X}, \lclock)$.
Our system-level objective is $\clockref \lclock. \; \eventop \globop \left(\costfn(x) = 0\right)$, i.e., our cost will eventually remain at zero.

\subsection{Specifying the system}

We consider the contracts that we will assign to each control component in our system.

\subsubsection{MPC}

We consider the specification of the MPC algorithm running on a dedicated processor with clock $\mclock$.
The MPC block will have an input $\hat x$, the state estimate coming from another control block.
It will have a single output $x_d$, which is a trajectory that the low-level controller has to follow.

On a given tick of the clock $\mclock$, the MPC block will assume that the state estimate is an accurate approximation of the state.
The MPC algorithm uses this to compute a trajectory in the next tick of $\mclock$.
Finally, the MPC algorithm will guarantee that the trajectories it generates
either decrease the cost $\costfn$ or keep it equal to zero.
We can represent the MPC contract $\cont^{\text{MPC}} = (A^{\text{MPC}}, G^{\text{MPC}})$ as follows:

{\footnotesize
\begin{align*}
    &A^{\text{MPC}}\colon \phi_{\text{A\_init}}^{\text{MPC}} \land \phi_{\text{timing}}^{\text{MPC}} \land \phi_{\text{sensor}}^{\text{MPC}}
    \land \phi_{\text{bound\_var}}^{\text{MPC}}\\
    &\quad\begin{aligned}
    &\phi_{\text{A\_init}}^{\text{MPC}} \colon \clockref \mclock.\;  \closepred\left(x, x_i; \delta_{\text{A\_init}}^{\text{MPC}}\right) \\
    &\phi_{\text{timing}}^{\text{MPC}} \colon \clockref \mclock.\;  \globop \left( \begin{aligned} & \left(T_{\text{min}}^\mclock \le \rclock(1) - \rclock \le T_{\text{max}}^\mclock\right) \land \\ &  \left(\rclock - \lclock^\rclock < T_{\text{fresh}}^\mclock\right) \end{aligned} \right) \\
    &\phi_{\text{sensor}}^{\text{MPC}} \colon \clockref \mclock.\;  \globop \closepred\left(\hat x, x; \delta_{\text{sensor}}^{\text{MPC}} \right)
    \\
    &\phi_{\text{bound\_var}}^{\text{MPC}} \colon \clockref \mclock.\;  \globop \texttt{BoundedVariation}(x; D_x)
    \end{aligned}
    \\
    &G^{\text{MPC}}\colon 
    \phi_{\text{G\_init}}^{\text{MPC}} \land \phi_{\text{rsp\_dyn}}^{\text{MPC}} \land \phi_{\text{dynamics}}^{\text{MPC}} \land \phi_{\text{traj\_bound\_var}}^{\text{MPC}} \land \phi_{\text{progress}}^{\text{MPC}} \\
    & \quad \begin{aligned}
    &\phi_{\text{G\_init}}^{\text{MPC}} \colon \clockref \mclock.\; \closepred\left(x_d(0)(0), x_i; \delta_{\text{G\_init}}^{\text{MPC}} \right)  \\
    &\phi_{\text{rsp\_dyn}}^{\text{MPC}} \colon \clockref \mclock.\;  \globop \texttt{RespectDynamics}(x_d) \\
    &\phi_{\text{dynamics}}^{\text{MPC}} \colon \clockref \mclock.\;  \globop \closepred\left(x_d(0)(T_{\text{avg}}^\mclock), x_d(1)(0); \delta_{\text{dynamics}}^{\text{MPC}}\right)\\
    &\phi_{\text{traj\_bound\_var}}^{\text{MPC}} \colon \clockref \mclock.\;  \globop \texttt{BoundedVariation}(x_d; D_d)  \\
    &\phi_{\text{progress}}^{\text{MPC}} \colon \clockref \mclock.\; \globop \left(\begin{aligned}
        &
        \left( \costfn\left(x_d(0)(0) \right) > 
            \costfn\left(x_d(1)(0) \right) \right) \lor \\ &
        \globop\left(\costfn\left(\texttt{Inflate}\left(\texttt{Im}\left(x_d\right) ; \delta_{\text{progress}}^{\text{MPC}} \right)\right) = 0 \right)
        \end{aligned}
        \right)
    \end{aligned}
\end{align*}}%
Assumption $\Phi_{\text{sensor}}^{\text{MPC}}$ captures the requirement that the sensor produces values that are close to the real state $x$ at the time when $\mclock$ ticks. The predicate $\closepred(v,v';\delta)$ indicates that two symbols $v, v'$ are evaluated to quantities that are close up to a parameter $\delta$ for some notion of distance (taken to be $\ell_2$ for the running example).
$\Phi_{\text{bound_var}}^{\text{MPC}}$ is an assumption that the state $x$ has bounded variation.
$\Phi_{\text{timing}}^{\text{MPC}}$
requires the clock period of $\mclock$ to lie between $T_{\min}^\mclock$ and $T_{\max}^\mclock$; (we also assume that the nominal $\mclock$ period $T_{\text{avg}}^\mclock$ lies between these bounds); the assumption also requires the difference in physical time between a tick of $\mclock$ and the tick of $\lclock$ from which $\mclock$ reads $\lclock$'s data to be less than $T_{\text{fresh}}^\mclock$. This means that the data from $\lclock$ that is read from $\mclock$ cannot be too old.
$\Phi_{\text{A\_init}}^{\text{MPC}}$ requires $x$ to be close to a value $x_i$ at the beginning of the system execution.

With respect to guarantees, $\Phi_{\text{G\_init}}^{\text{MPC}}$ 
means that the first trajectory output by the MPC block will start close to $x_i$.
$\Phi_{\text{rsp\_dyn}}^{\text{MPC}}$ is a predicate stating that MPC will always generate trajectories $x_d$ that should be within the competence of the low-level block to follow. In the context of this work, we require that the produced trajectories are dynamically feasible, i.e., there exists a feedback controller which is able to track the generated trajectories.
$\Phi_{\text{dynamics}}^{\text{MPC}}$ states that the beginning of every new trajectory provided by the MPC block has to be close to the value of the previous trajectory at time $T_{\text{avg}}^\mclock$.
$\Phi_{\text{traj_bound_var}}^{\text{MPC}}$ is a promise that the trajectories provided by MPC will have bounded variation.

Finally, in order to be able to promise that the system will make progress towards its objective, the MPC block will make use of the cost function $\costfn$.
The MPC block makes a guarantee $\Phi_{\text{progress}}^{\text{MPC}}$ which says that either 
the cost at the beginning of a trajectory $x_d$ is larger than the cost at the beginning of the next trajectory $x_d$, or that all points that are close to the trajectory have a cost of zero. The function $\texttt{Inflate}(A,\delta)$ takes a set $A$ and returns the set of all points that are $\delta$-close to any point of $A$.
In other words, the promise is that each trajectory improves the cost or stays fixed at cost equal to zero.

\subsubsection{Low-level controller}

The low-level controller takes as inputs trajectories $x_d$ that the system's state $x$ has to follow and promises that it can make the system follow these trajectories with a given accuracy. We have the following contract $\cont^{\text{FL}} = (A^{\text{FL}}, G^{\text{FL}})$ for the low-level controller:

{\footnotesize
\begin{align*}
    A^{\text{FL}}\colon & \phi_{\text{timing}}^{\text{FL}} \land \phi_{\text{rsp\_dyn}}^{\text{FL}} \land \phi_{\text{dynamics}}^{\text{FL}} \land \phi_{\text{bound\_var}}^{\text{FL}}\\
    &\phi_{\text{timing}}^{\text{FL}} \colon \clockref \lclock.\; \globop \left( \begin{aligned} & \left(T_{\text{min}}^\lclock \le \rclock(1) - \rclock \le T_{\text{max}}^\lclock\right) \land \\ & \left(\rclock - \mclock^\rclock < T_{\text{fresh}}^\lclock\right) \end{aligned} \right) \\
    &\phi_{\text{rsp\_dyn}}^{\text{FL}} \colon \clockref \lclock.\; \globop \texttt{RespectDynamics}(x_d) \\
    &\phi_{\text{dynamics}}^{\text{FL}} \colon \clockref \lclock.\; 
    \globop \left( \begin{aligned} & (\mclock \ne \mclock(-1)) \land (\mclock \ge 0) \\ & \to \closepred\left(x_d(0)(0), x; \delta_{\text{dynamics}}^{\text{FL}}\right) \end{aligned} \right)\\
    &\phi_{\text{bound\_var}}^{\text{FL}} \colon \clockref \lclock.\; \globop \texttt{BoundedVariation}(x; D_x)\\
    G^{\text{FL}}\colon & \phi_{\text{upd}}^{\text{FL}} \land \phi_{\text{tracking}}^{\text{FL}} \\
    &\phi_{\text{upd}}^{\text{FL}} \colon \clockref \lclock.\; \globop \left( \begin{aligned} & \textbf{if}\;\; (\mclock \ne \mclock(-1)) \land (\mclock \ge 0) \;\; \textbf{then} \\ & (\upd = \lclock) \;\;\textbf{else} \;\; (\upd = \upd(-1)) \end{aligned} \right)\\
    &\phi_{\text{tracking}}^{\text{FL}} \colon \clockref \lclock.\; \globop \left(
    \begin{aligned}
    &
    (\lclock - \upd > 0) \to \\
    &
    \closepred\left(x, x_d(0)\left( T_{\text{avg}}^\lclock (\lclock - \upd) \right); \delta_{\text{tracking}}^{\text{FL}}\right)
    \end{aligned}
    \right)
\end{align*}}
The assumptions of this contract are as follows.
$\phi_{\text{timing}}^{\text{FL}}$ is similar to $\phi_{\text{timing}}^{\text{MPC}}$: it assumes that $\lclock$'s period is bounded below and above, and that the data read from clock $\mclock$ is not too old.
$\phi_{\text{rsp\_dyn}}^{\text{FL}}$ requires trajectories received from $\mclock$ to respect certain physical limits, i.e., that they satisfy the state and input constraints used in the analysis of the low level controller.
$\phi_{\text{dynamics}}^{\text{FL}}$ makes sure that when $\lclock$ detects the generation of a new trajectory $x_d$ from $\mclock$, then the starting point of that trajectory should be close to the value of the state.
$\phi_{\text{bound_var}}^{\text{FL}}$ requires the state $x$ to have bounded variation.

Regarding guarantees, $\phi_{\text{upd}}^{\text{FL}}$ is a helper statement that defines the variable $\texttt{upd}$. This variable contains the last value of $\lclock$ when a new trajectory was received from $\mclock$.
$\phi_{\text{tracking}}^{\text{FL}}$ guarantees that the low-level controller will make the system follow the given trajectory $x_d$. Observe that $\phi_{\text{tracking}}^{\text{FL}}$ is enforced for all values of the trajectory $x_d$, except its first point. For the first point of the trajectory, the low level controller makes the assumption $\phi_{\text{dynamics}}^{\text{FL}}$ on the state.
$T_\text{avg}^\lclock$ is the nominal period of $\lclock$ and respects the bounds of $\phi_{\text{timing}}^{\text{MPC}}$.

\subsubsection{Estimator}
The estimator will guarantee that the state estimates are always accurate on the clock ticks of $\lclock$, yielding the contract $\cont^{\text{Est}} = (A^{\text{Est}}, G^{\text{Est}})$
{\footnotesize
\begin{align*}
    A^{\text{Est}}\colon & \clockref \lclock.\; \texttt{True}\\
    G^{\text{Est}}\colon & \phi_{\text{sensor}}^{\text{Est}} \\
    &\phi_{\text{sensor}}^{\text{Est}} \colon \clockref \lclock.\; \globop \closepred\left(\hat x, x; \delta_{\text{sensor}}^{\text{Est}}\right)
\end{align*}}

\subsubsection{Timing design}
We understand the timing component of the system as ``network design'' in the sense that
enforcing timing constraints is implemented by applying networking and clock synchronization technologies. This component will have the following contract $\cont^{\text{Tmg}} = (A^{\text{Tmg}}, G^{\text{Tmg}})$:
{\footnotesize
\begin{align*}
    A^{\text{Tmg}}\colon & \clockref \rclock.\; \texttt{True}\\
    G^{\text{Tmg}}\colon & \phi_{\lclock\text{-timing}}^{\text{Tmg}}\land \phi_{\mclock\text{-timing}}^{\text{Tmg}} \\
    &\phi_{\lclock\text{-timing}}^{\text{Tmg}} \colon \clockref \lclock.\; \globop \left( \begin{aligned} & \left(T_{\text{min}}^\lclock \le \rclock(1) - \rclock \le T_{\text{max}}^\lclock\right) \land \\ &  \left(\rclock - \mclock^\rclock < T_{\text{fresh}}^\lclock\right) \end{aligned} \right) \\
    &\phi_{\mclock\text{-timing}}^{\text{Tmg}} \colon  \clockref \mclock.\; \globop \left( \begin{aligned} & \left(T_{\text{min}}^\mclock \le \rclock(1) - \rclock \le T_{\text{max}}^\mclock\right) \land \\ &  \left(\rclock - \lclock^\rclock < T_{\text{fresh}}^\mclock\right) \end{aligned} \right)
\end{align*}}

\subsection{System-level analysis}
\label{sc:system-analysis}

\label{sec:system-analysis}
\begin{figure}
    \centering
    \includegraphics[width=\columnwidth]{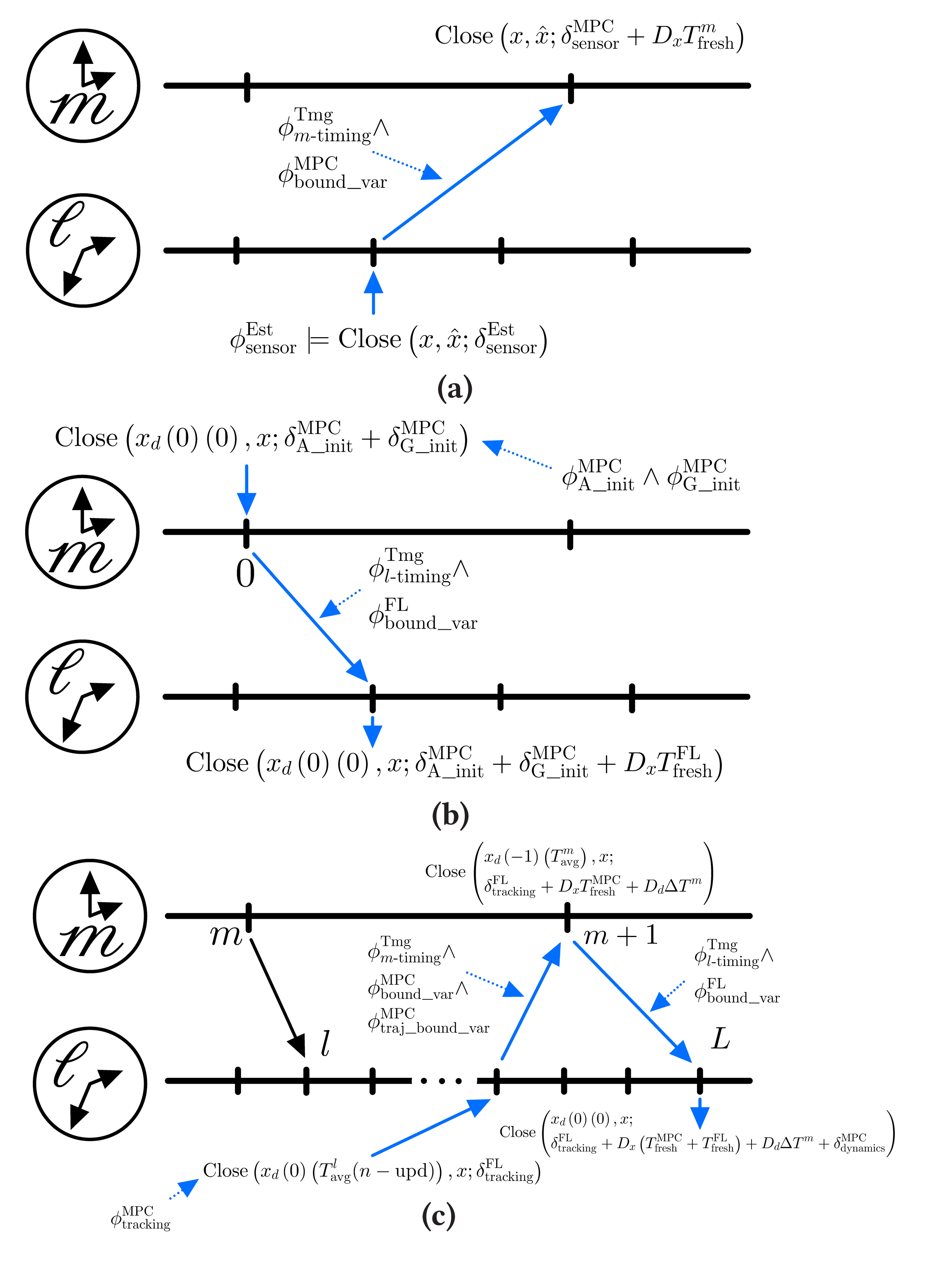}
    \vspace{-10mm}
    \caption{Timing diagrams showing how to satisfy the assumptions (a) $\phi_{\text{sensor}}^{\text{MPC}}$, (b) $\phi_{\text{dynamics}}^{\text{FL}}$ (base step), and (c) $\phi_{\text{dynamics}}^{\text{FL}}$ (inductive case)}
    \label{kjqkjbfkjqhg}
    \vspace{-5mm}
\end{figure}

At this point, we have component-level contracts for all elements in our system. The contract operation of composition yields the specification formed by interconnecting the components whose specifications we have available.
Thus, our system-level specification is
\begin{equation}\label{kghckjsgb}
\cont^{\text{Sys}} = \cont^{\text{MPC}} \parallel \cont^{\text{FL}} \parallel \cont^{\text{Est}} \parallel \cont^{\text{Tmg}}.
\end{equation}
As pointed out in \cite{pactipaper}, any component is a valid implementation of a relaxation of its specification (i.e., of any contract $\cont'$ such that $\cont \le \cont'$, where $\cont$ is the component's contract). The authors propose an algorithm to compute abstractions of contract compositions that ensure that all components in the system operate with their assumptions satisfied. We will apply the \texttt{ContractComposition} routine of Algorithm~1 of \cite{pactipaper} in order to compute the composition $\cont^{\text{Sys}}$. We consider how the composition of components yields a system satisfying the assumptions of all components.

\subsubsection{MPC} We consider the assumptions of MPC:
$\phi_{\text{init}}^{\text{MPC}}$ is an assumption on the initial state that we need to conserve at the system-level. $\phi_{\text{bound\_var}}^{\text{MPC}}$ is also a system-level assumption.
$\phi_{\text{timing}}^{\text{MPC}}$ is satisfied by $\phi_{\mclock\text{-timing}}^{\text{Tmg}}$ of our timing design.
$\phi_{\text{sensor}}^{\text{MPC}}$ needs to be analyzed.

We observe that
{\footnotesize
$$
\infer{\clockref \mclock.\; \closepred\left( \hat x, x; \delta_{\text{sensor}}^{\text{Est}} + T_{\text{fresh} }^{\mclock} D_x \right)}
{
\phi_{\text{sensor}}^{\text{Est}} & \phi_{\mclock\text{-timing}}^{\text{Tmg}} & \phi_{\text{bound\_var}}^{\text{MPC}}
}.
$$}%
This notation means that the formula below the horizontal bar is a valid deduction from the conjunction of the formulas on top of the bar. We need the following relation among the parameters in order to satisfy $\phi_{\text{sensor}}^{\text{MPC}}$:
\mybox{
\begin{equation}\label{jhgbfkjk}
    \delta_{\text{sensor}}^{\text{Est}} + T_{\text{fresh} }^{\mclock} D_x \le \delta_{\text{sensor}}^{\text{MPC}}.
\end{equation}}
Figure~\ref{kjqkjbfkjqhg}a represents this inference in a timing diagram.

\subsubsection{Low-level controller}
Now we verify the assumptions of the low-level controller.
$\phi_{\text{timing}}^{\text{FL}}$ is satisfied by $\phi_{\lclock\text{-timing}}^{\text{Tmg}}$ of the timing design.
$\phi_{\text{bound\_var}}^{\text{FL}}$ is an assumption on the dynamics of the physical system and should therefore be a system-level assumption.
$\phi_{\text{rsp\_dyn}}^{\text{FL}}$ is satisfied by the guarantee $\phi_{\text{rsp\_dyn}}^{\text{MPC}}$.
The satisfaction of 
$\phi_{\text{dynamics}}^{\text{FL}}$ requires an inductive argument.

The first time that the antecedent in $\phi_{\text{dynamics}}^{\text{FL}}$ holds, we have
{\footnotesize
$$
\infer{
    \clockref \lclock.\; 
    \globop \left( \begin{aligned} & (\mclock \ne \mclock(-1)) \land (\mclock = 0) \\ & \to \closepred\left(x_d(0)(0), x; \delta_{\text{G\_init}}^{\text{MPC}} + \delta_{\text{A\_init}}^{\text{MPC}} + D_x T_{\text{fresh}}^\lclock \right) \end{aligned} \right)
}{
\infer{
    \clockref \mclock.\; \closepred\left(x_d(0)(0), x; \delta_{\text{G\_init}}^{\text{MPC}} + \delta_{\text{A\_init}}^{\text{MPC}} \right)
}
{\phi_{\text{A\_init}}^{\text{MPC}} & \phi_{\text{G\_init}}^{\text{MPC}}}
& \phi_{\text{bound\_var}}^{\text{FL} } & \phi_{\text{timing}}^{\text{FL}}
}.
$$}%
Therefore, we must have
\mybox{
\begin{equation}\label{jhgkjgkjdh}
\delta_{\text{G\_init}}^{\text{MPC}} + \delta_{\text{A\_init}}^{\text{MPC}} + D_x T_{\text{fresh}}^\lclock \le \delta_{\text{dynamics}}^{\text{FL}}.
\end{equation}}
\noindent
in order for $\phi_{\text{dynamics}}^{\text{FL}}$ to hold in the initial case---see Figure~\ref{kjqkjbfkjqhg}b.
Now we inductively assume that $\phi_{\text{dynamics}}^{\text{FL}}$ holds at $l \in \lclock$.
We analyze the case when the antecedent is true. We will verify the conditions required so that the predicate holds the next time the antecedent is true.
Let $m = \tau_\lclock^\mclock(l)$
and $L = \min \setArg{x \in \lclock}{\tau_\lclock^\mclock = m + 1}$. We have the following inference:
{\tiny
$$
\infer{
    \clockref \lclock.\; \globop_{[L,L]} \closepred\left( 
    x, x_d(0)(0);\;
    \delta_{\text{tracking}}^{\text{FL}} + \delta_{\text{dynamics}}^{\text{MPC}} + \left(T_{\text{fresh}}^\mclock + T_{\text{fresh}}^\lclock\right) D_x + D_d \Delta T^\mclock
    \right)
}{
\infer{
    \clockref \mclock.\; \globop_{[m+1,m+1]} \closepred\left( 
    x, x_d(0)(0);\;
    \delta_{\text{tracking}}^{\text{FL}} + \delta_{\text{dynamics}}^{\text{MPC}} + T_{\text{fresh}}^\mclock D_x + D_d \Delta T^\mclock
    \right)
}
{
\infer{
    \clockref \mclock.\; \globop_{[m+1,m+1]} \closepred\left( 
    x, x_d(-1)(T_{\text{avg}}^\mclock);\;
    \delta_{\text{tracking}}^{\text{FL}} + T_{\text{fresh}}^\mclock D_x + D_d \Delta T^\mclock
    \right)
}
{
    \phi_{\text{tracking}}^{\text{FL}} & \phi_{\text{timing}}^{\text{MPC}} & \phi_{\text{bound\_var}}^{\text{MPC}} & \phi_{\text{traj\_bound\_var}}^{\text{MPC}} 
}
& \phi_{\text{dynamics}}^{\text{MPC}}
}
&
\phi_{\text{timing}}^{\text{FL}}
},
$$}%
where $\Delta T^\mclock = T_{\text{max}}^\mclock - T_{\text{avg}}^\lclock \left\lfloor \frac{T_{\text{min}}^\mclock - \left(T_{\text{fresh}}^\mclock + T_{\text{fresh}}^\lclock\right)}{T_{\max}^\lclock} \right\rfloor$
captures the maximum physical time difference between a period of clock $\mclock$ and the amount of physical time that clock $\lclock$ can account as having elapsed since it read a new trajectory $x_d$ from MPC. From the perspective of clock $\lclock$, each of its ticks is separated by an interval of $T_{\text{avg}}^\lclock$ time units. From our analysis, if the expression
\mybox{
\begin{equation}\label{bfhqglbgk}
    \delta_{\text{tracking}}^{\text{FL}} + \delta_{\text{dynamics}}^{\text{MPC}} + (T_{\text{fresh}}^\mclock + T_{\text{fresh}}^\lclock) D_x + D_d \Delta T^\mclock \le \delta_{\text{dynamics}}^{\text{FL}}
\end{equation}}
\noindent
is true, $\phi_{\text{dynamics}}^{\text{FL}}$ will hold inductively. This analysis is shown graphically in Figure~\ref{kjqkjbfkjqhg}c.

\subsubsection{Progress}
Our analysis indicates that the composition of all contracts so far defined yields a situation in which all contracts have their assumptions met---provided the system parameters meet conditions \eqref{jhgbfkjk}, \eqref{jhgkjgkjdh}, and \eqref{bfhqglbgk}. Now we verify whether the system makes progress towards its goal.
We observe that the fact that $\costfn$ takes values in a well order means that we can carry out the following deductions:
{\footnotesize
$$
\infer{\clockref \lclock.\; \eventop\globop \left(\costfn(x) = 0\right)}
{
\infer{\clockref \mclock.\; \eventop\globop\left(\costfn\left(\texttt{Inflate}\left(\texttt{Im}\left(x_d\right) ; \delta_{\text{progress}}^{\text{MPC}} \right)\right) = 0 \right)}
{\phi_{\text{progress}}^{\text{MPC}}}
&
\phi_{\text{upd}}^{\text{FL}}
&
\phi_{\text{tracking}}^{\text{FL}}
&
\phi_{\text{dynamics}}^{\text{FL}}
}
$$}%
provided that
\mybox{
\begin{equation}\label{kjgbqjkg}
\delta_{\text{dynamics}}^{\text{FL}} \le \delta_{\text{progress}}^{\text{MPC}}.
\end{equation}}

\subsubsection{System-level contract}
From our analysis so far, provided that conditions \eqref{jhgbfkjk}, \eqref{jhgkjgkjdh}, \eqref{bfhqglbgk}, and \eqref{kjgbqjkg} hold, the relaxation of contract $\cont^{\text{Sys}}$ given by \eqref{kghckjsgb} and computed using Algorithm 1 of \cite{pactipaper} is
\begin{equation}
\label{kjhbxgkjdg}
\begin{split}
    A\colon &\phi_{\text{A\_init}}^{\text{MPC}} \land \phi_{\text{bound\_var}}^{\text{MPC}}\\
    G\colon &
    \clockref \lclock.\; \eventop\globop \left(\costfn(x) = 0\right),
\end{split}%
\end{equation}%
\noindent
which means that our system satisfies the desired top-level stability objective.

\subsection{Component-level verifications}
\label{jbhkxnlbbg}

Our system-level analysis shows that if we implement control blocks adhering to the specifications discussed in Section~\ref{sc:system-analysis}, our system will satisfy the desired stability property---see \eqref{kjhbxgkjdg}.
Now we have to verify that each implementation of our control blocks satisfies its own component-level contract.

\subsubsection{Low-level controller}
For the running example, we will employ the control strategy of C-MPC---a B\'ezier curve MPC variant---combined with a feedback linearization low level controller, as developed in \cite{csomay-shanklin-multi-rate-2022}. This will allow us to make continuous time guarantees with discrete time constraints.
In order to use the methods presented therein, we begin by showing that applying zero order held inputs, i.e., with $\clockref \lclock.\; \globop \left(
    \bar{u} = k_\text{fbl}(x, \lclock - \texttt{upd})
    \right)$
results in bounded exogenous disturbance to the error dynamics. For the following discussion, let $x'(t)$ denote the solution to the system dynamics with continuous time control applied, and $x(t)$ the solution with zero order held inputs. Plugging in the solution to the differential equation with $x'(0) = x(0)$ and adding and subtracting $g(x'(\tau))\bar{u}$ yields

{\footnotesize
\begin{align}
    \|x(t) - x'(t)\| = \int_0^t&\frac{d}{dt}(x(\tau) - x'(\tau)) d\tau \label{eq:der-bound}\\ 
    \le \int_0^t&\|f(x(\tau)) + g(x(\tau))\bar{u}-g(x'(\tau))\bar{u} - f(x'(\tau))  \notag \\
    &- g(x'(\tau))u(x'(\tau),\tau) + g(x'(\tau))\bar{u}\|d\tau \notag \\
    \le \int_0^t&(L_f +L_g\|\bar{u}\|) \|x(\tau) - x'(\tau)\| \notag \\
    & \ \ \ \ \ + \|g(x'(\tau))\|\|u(\tau) - \bar{u}\|d\tau, \notag 
\end{align}}%
where $L_f$ and $L_g$ represent the local Lipschitz constants of the drift vector and actuation matrix, respectively. From the perspective of the low level controller, we assume via $\phi^\text{FL}_\text{rsp$\_$dyn}$ that $\| u(t)\|\le U$ and that $x(t) \in \mathcal{X}$, a compact set. These will be necessary for the analysis, and will be explicitly enforced in the proposed MPC formulation.
From these constraints, we know that there exists a $G>0$ such that $\|g(x'(\tau))\|<G$ and that all local Lipschitz constants are global over $\mathcal{X}$. Combining these facts and using the Bellman-Gronwall Lemma \cite{sastry-mathematical-1999} leads to
\begin{align*}
    \|x(t) - x'(t)\| \le {UG}t^2e^{(L_f+L_gU)t} \triangleq t\rho(t),
\end{align*}
where $\rho\in \mathcal{K}_\infty$, a class $\mathcal{K}$ infinity function. Next, let $e'(t) \triangleq x'(t) - x_d(t)$ denote the error dynamics with continuous time control applied, whereby the feedback linearizing controller in \eqref{eq:fbl} yields $\dot{e}'(t) = A_\text{cl}e'$ for $A_\text{cl}$ a stable matrix. Taking $e(t) = x(t) - x_d(t)$ to be the error with zero order held inputs, we have
\begin{align*}
    \dot e = \underbrace{\frac{d}{dt}(x - x') + A_\text{cl}(x'-x)}_{\triangleq w(t)} + A_\text{cl}e.
\end{align*}
Plugging in the terms developed in \eqref{eq:der-bound} results in
\begin{align*}
    \|w(t)\| \le \|w(T^\lclock_\text{max})\| \le T^\lclock_\text{max}((L_f + 2L_gU + \|A_\text{cl}\|)\rho(T^\lclock_\text{max}) + G U).
\end{align*}
As this bound is a composition of class-$\mathcal{K}$ functions in time, for all $\delta_\text{w} > 0$ there exists an $\epsilon > 0$ such that $T^\lclock_{\text{max}} < \epsilon$ results in $\|w(t)\| \le \delta_\text{w}$. Fixing an allowable $\delta_\text{w}$ and thereby upper bounding $T^\lclock_\text{max}$, by integrating the error dynamics and using the comparison lemma we have
\begin{align*}
    \|e(t) \| \le \| x(0) - x_d(0)\| Me^{-\lambda t} + T^\lclock_\text{max}\delta_\text{w},
\end{align*}
for some $M, \lambda > 0$ as determined by the convergence rate of the low level controller. 
Therefore, in order to produce the guarantee of $\phi^\text{FL}_\text{tracking}$, we must enforcing that 
\begin{align}
   \delta^\text{FL}_\text{dynamics} Me^{-\lambda T^\lclock_\text{min}} + T^\lclock_\text{max} \delta_\text{w} \le \delta^\text{FL}_\text{tracking}.
\end{align}


\subsubsection{High-level controller}
For the sake of exposition, we condense the C-MPC program presented in \cite{csomay-shanklin-multi-rate-2022} into what it enforces:

{\footnotesize
\begin{equation}
\begin{aligned}
\label{eq:MPC}
\min_{x_d(t), u_d(t)} \quad & \int_0^T h(x_d(t), u_d(t)) + J(x_d(T)) \\
\textrm{s.t.} \quad & x_d(t) =f^a(x_d(t)) + g^a(x_d(t))u_d(t), \\
  &x_d(0) \in \hat{x} \oplus\mathcal{E},~~~~~\ \ \ x_d(T) = 0, \\
  &x(t) \in \mathcal{X} ,\ \ \ \  \ \ \ \ \ \ \ u(t) \in \mathcal{U}. \\
\end{aligned}
\end{equation}}%
with $h\colon \mathcal{X}\times\mathcal{U} \to \mathbb{R}_{\ge 0}$ a convex stage cost, $J\colon \mathcal{X}\to\mathbb{R}_{\ge 0}$ a convex terminal cost, $f^a$ and $g^a$ linear approximations of the dynamics, $\mathcal{E}\subset\mathbb{R}^n$ the assumed robust invariant set, and $x(t)$ and $u(t)$ the continuous-time state and input of the low-level system. Lemma 2 in \cite{csomay-shanklin-multi-rate-2022} demonstrates that the trajectories produced are dynamically feasible, i.e., able to be exactly tracked via the feedback linearizing controller in \eqref{eq:fbl}, implying satisfaction of $\phi_\text{rsp$\_$dyn}^\text{MPC}$. As the MPC program produces solutions to a linear system with bounded state and control inputs, we have that the resulting desired trajectory $x_d$ will have bounded variation, satisfying $\phi_\text{traj$\_$bound$\_$var}^\text{MPC}$.

Next, must ensure that the above MPC program is recursively feasible, which requires showing that the enforced set $\mathcal{E}$ is a robust invariant for the system. Choosing $\mathcal{E}$ such that $B\left(0, \delta_\text{tracking}^\text{FL}\right) \subset \mathcal{E}$ results in a set that can be rendered invariant by the low level controller, even during sampling. Therefore we are free to choose $\mathcal{E}$ to meet this condition as well as the design requirements imposed by the system analysis in Section~\ref{sec:system-analysis}. With this, we can appeal to the results in \cite{csomay-shanklin-multi-rate-2022} to prove recursive feasibility of the MPC algorithm used therein.

Finally, let $V\colon \mathcal{X}\to\mathbb{R}$ denote the sum of the running and terminal cost of MPC, often used as a Lyapunov function in stability proofs. As we have effectively transformed our nonlinear MPC program to a linear one, we can use standard MPC results \cite{borrelli-predictive-nodate} to state that $\clockref \mclock.\; \left( V(x_d(1)(0)) < V(x_d(0)(0)) \right)$.
Since the guarantees of the MPC contract involve $\costfn$, which takes values in a well-order,
we can define $\costfn$ by quantizing the function $V$. 
Then we can choose a vicinity around $x_g$ and define $\costfn (p) = 0$ for $p \in \mathcal{E}$. In that case, $\costfn$ takes values in a well-order, and the MPC block satisfies $\phi_{\text{progress}}^{\text{MPC}}$.


%
%
%
%

\subsection{Simulation results}
\begin{figure}[t!]
    \centering
    \includegraphics[width=0.95\columnwidth]{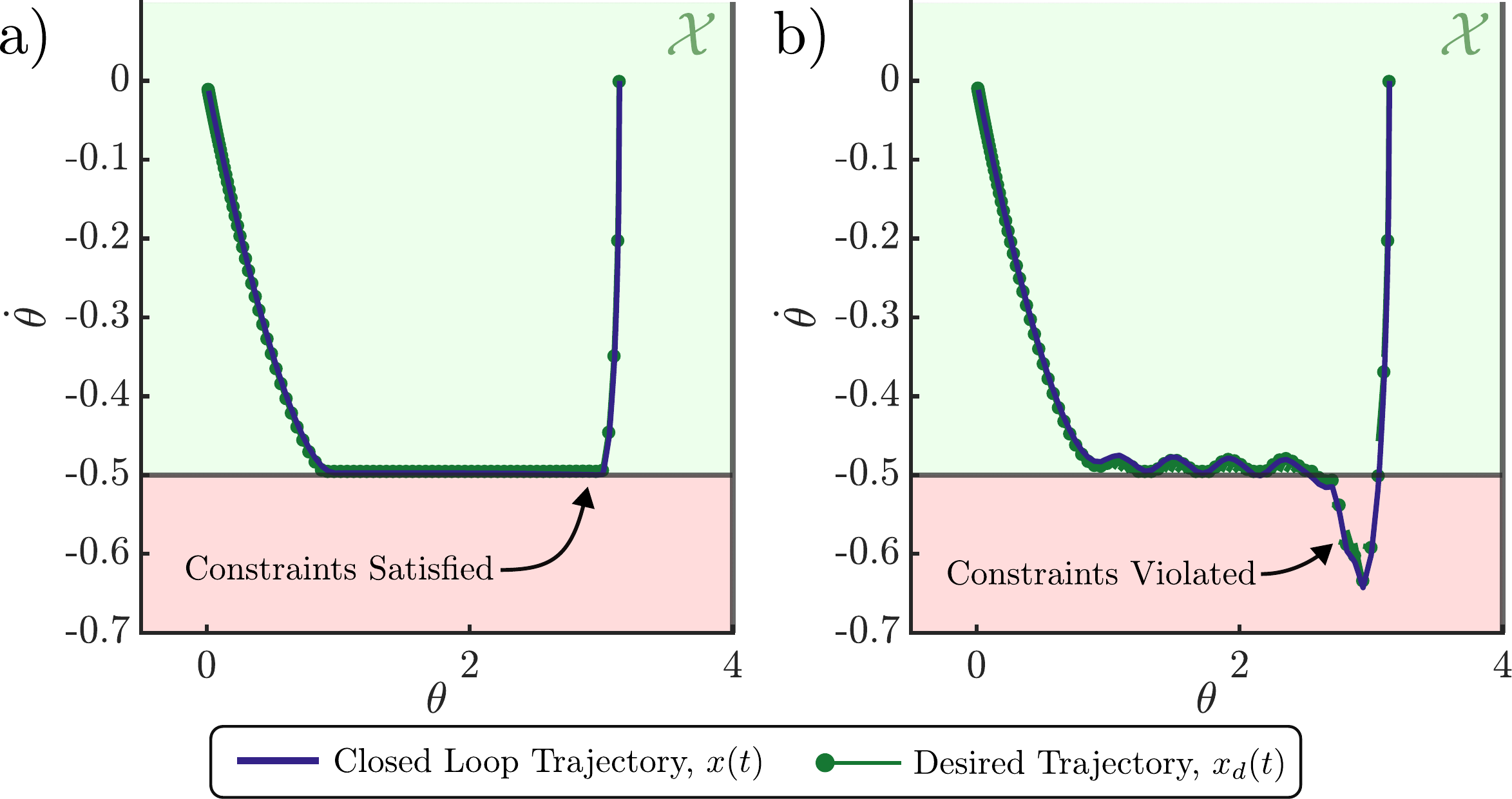}
    \caption{Increasing time delay beyond $T^\lclock_\text{fresh}$ violates component assumption $\phi^\text{FL}_\text{timing}$, breaking expression \eqref{bfhqglbgk} and leading to closed-loop constraint violation.}
    \label{fig:simulation-results}
    \vspace{-4mm}
\end{figure}
As seen in Figure~\ref{fig:simulation-results}a, when all of the defined specifications are met, we are able to satisfy the desired state constraint, despite the sampling and holding of low level control inputs. In Figure~\ref{fig:simulation-results}b, we simulate the effect of a time delay of duration $2T^\text{MPC}_\text{min}$ in sending the computed desired trajectory $x_d$ from the MPC layer to the low level controller. A time delay of this form is always present in practice, as the optimal control problem takes a nontrivial amount of time to compute, but is often ignored. This time delay violates constraint \eqref{bfhqglbgk}, which is required for our system-level analysis of Section~\ref{sec:system-analysis} to hold, and results in violation of the enforced state constraint for the closed loop control system.

\section{Concluding remarks}
\label{sc:conclusion}

This paper considered the use of assume-guarantee specifications in the analysis of networked/layered control systems operating on multiple clocks. In order to express the specifications for control components that operate on their own time base, we introduced the syntax and semantics of the logic \lname. We showed that we could rigorously express specifications that are intuitive to designers. We considered how to independently specify the MPC and feedback linearization blocks used to stabilize a pendulum.
By operating on the component contracts, we showed that a system implemented by interconnecting the control blocks in our system yields a control system that meets our objectives. Then we showed how to verify that each control block satisfies its own specification. Our simulation results showed that a violation of the system-level constraints may lead to failure.


\bibliographystyle{style/ACM-Reference-Format}
\balance
\bibliography{support/references}

\end{document}